\documentclass[manuscript,screen]{acmart}
\AtBeginDocument{%
  \providecommand\BibTeX{{%
    \normalfont B\kern-0.5em{\scshape i\kern-0.25em b}\kern-0.8em\TeX}}}

\setcopyright{acmlicensed}
\copyrightyear{2018}
\acmYear{2018}
\acmDOI{XXXXXXX.XXXXXXX}

\acmConference[CHI 2024]{Sensemaking Workshop}{May 11--16,
  2024}{Honolulu, Hawai'i}
\acmBooktitle{Woodstock '18: ACM Symposium on Neural Gaze Detection,
 June 03--05, 2018, Woodstock, NY} 
\acmISBN{978-1-4503-XXXX-X/18/06}

\begin{document}

\title{What to Make Sense of in the Era of LLM? A Perspective from the Structure and Efforts in Sensemaking}

\author{Tianyi Li}
\email{li4251@purdue.edu}
\affiliation{%
  \institution{Purdue University}
  \city{West Lafayette}
  \state{Indiana}
  \country{USA}
}
\author{Satya Samhita Bonepalli}
\affiliation{%
 \institution{Purdue University}
  \city{West Lafayette}
  \state{Indiana}
  \country{USA}
}
\email{sbonepal@purdue.edu}
\author{Vikram Mohanty}
\email{vikrammohanty@acm.org}
\affiliation{%
  \institution{Bosch Research North America}
  \city{Sunnyvale}
  \state{California}
  \country{USA}
}







\renewcommand{\shortauthors}{Li et al.}

\begin{abstract}

  Sensemaking tasks often entail navigating through complex, ambiguous data to construct coherent insights. Prior work has shown that crowds can effectively distribute cognitive load, pooling diverse perspectives to enhance analytical depth. Recent advancements in LLMs have further expanded the toolkit for sensemaking, offering scalable data processing, complex pattern recognition, and the ability to infer and propose meaningful hypotheses. In this study, we explore how LLMs (i.e., GPT-4) can assist in a complex sensemaking task of deciphering fictional terrorist plots. We explore two different approaches for leveraging GPT-4's capabilities: a holistic sensemaking process and a step-by-step approach. Our preliminary investigations open the doors for future research into optimizing human-AI collaborative workflows, aiming to harness the complementary strengths of both for more effective sensemaking in complex scenarios.
\end{abstract}

\begin{teaserfigure}
    \centering
  \includegraphics[width=0.8\textwidth]{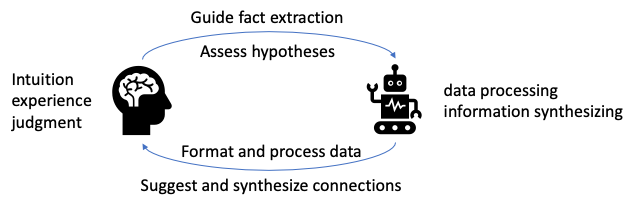}
  \caption{One possible human-LLM collaboration mechanisms in sensemaking tasks.}
  \Description{Enjoying the baseball game from the third-base
  seats. Ichiro Suzuki preparing to bat.}
  \label{fig:teaser}
\end{teaserfigure}


\maketitle

\section{Introduction}
Sensemaking is a sophisticated cognitive process that involves interactions with vast amounts of data that can be inherently complex and ambiguous. Prior research has facilitated this process through visual interactions with the dat~\cite{7042492} and enhance collaborative efforts via communication and handoff of intermediate results~\cite{zhao2017supporting} to enable more people to contribute to the sensemaking process. Additionally, methodologies have been developed to distribute the cognitive load across a wide array of novice participants~\cite{li2018crowdia}.
Recently, the advancement of Large Language Models (LLMs) has introduced revolutionary potential to address some of the challenges in sensemaking tasks, suggesting a significant shift in how we approach and manage these complex cognitive processes.

Meanwhile, LLMs are susceptible to generating hallucinations and inaccuracies, casting doubt on their reliability for solving complex sensemaking tasks, such as those involving crime mysteries. This position paper delves into the research question: \textbf{how can LLMs assist human experts in analyzing complex sensemaking tasks?} To investigate this, we examine their application to making sense of a dataset related to a fictional terrorist attack.

Through the framework of human expert sensemaking loop~\cite{pirolli2005sensemaking} proposed by Pirolli and Card, alongside the crowd sensemaking pipeline~\cite{li2018crowdia} modeled in their spirit, this paper explores the integration of LLMs into the human sensemaking process. It assesses the quality of analysis produced by LLMs and identifies new synergies between human and AI efforts in collaborative sensemaking. The study evaluates LLM-generated analysis with prompt design informed by instructions for experts and layperson, and compare these with analyses by novice crowd workers.

\section{Methods}
A preliminary investigation was conducted utilizing a dataset comprising 10 documents. The sensemaking task is to formulate several plausible hypotheses for identifying potential terrorist activities. 
This study examines the sensemaking capabilities of novice human analysts compared to LLMs. The performance of human analysts was gauged against a baseline established in previous research~\cite{10.1145/3359238}, involving crowdsourced sensemaking tasks executed by non-expert individuals sourced from Amazon Mechanical Turk. 
With LLMs, there is an infinite querying space. This paper explores two different strategies --- entrust LLMs to develop alternative hypotheses based on the dataset; and request LLMs to follow a predefined set of steps to produce a series of intermediate analysis. The first strategy is hereafter referred to as the ``\textbf{holistic}'' approach and the second as the ``\textbf{step-by-step}'' approach.

While evaluating the quality of sensemaking necessitates a nuanced examination of the validity of the procedures, potential analytical biases, the representativeness and pertinacity of the data utilized, and other factors, as an initial attempt, we focus on two initial metrics: the LLM's ability to extract relevant facts from the documents and its effectiveness in generating hypotheses that align with those specified in the answer key. 

\section{Considerations for Assessing Sensemaking Performance}
While the answer key provides a structured list of ``facts'' and their interconnections leading to various hypotheses, it is challenging to quantify how well the ``facts'' were extracted by different methods without a well defined unit. For example, a fact from the answer key is 
\begin{quote}
    ``Erica Hahn, of North Bergen, NJ, has deposited checks in his bank account that were drawn on account number <\#\#\#> held by Atticus Lincoln at the First Union Bank in Springfield, VA''.
\end{quote} And a fact extracted in one of the conditions was
\begin{quote}
    ``Erica Hahn has financial ties to Atticus Lincoln''.
\end{quote}
Although both versions convey the same fundamental information, the latter version omits several details. This level of abstraction does not impede the development of the final hypotheses, yet it lacks the analytical depth and rigor implied by the answer key. 

Therefore, we evaluate sensemaking effectiveness under the different conditions using two key metrics: (1) \textbf{Coverage}, which assesses the extent to which the analysis encompasses the relevant facts/hypotheses, and (2) \textbf{Comprehensiveness}, which evaluates the degree to which the details within the facts/hypotheses presented in the answer key are accurately captured and reflected in the analysis.

\section{Preliminary Results}
\begin{figure}
    \centering
    \includegraphics[width=0.75\linewidth]{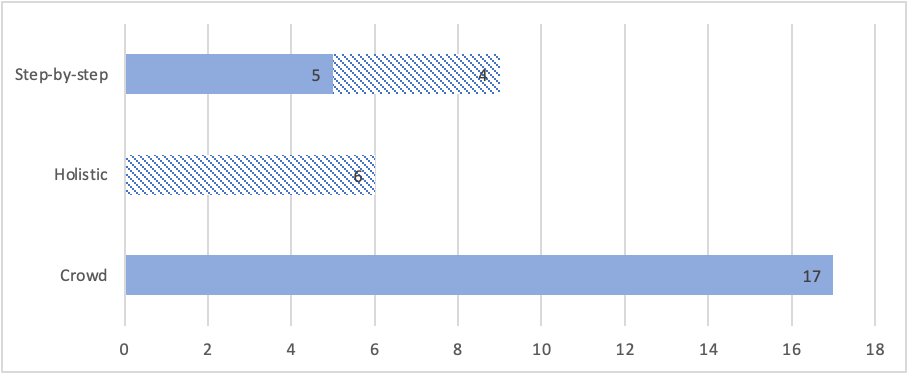}
    \caption{Efficacy in Fact Extraction (shadowed parts are facts that are not comprehensively covered).}
    \label{fig:facts}
\end{figure}

Regarding efficacy of \textit{fact extraction}, previous research findings indicate that \textbf{crowdsourced} sensemaking analysis successfully encompassed 17 of the key relevant facts, all are comprehensive. With the \textbf{``holistic''} approach, LLM sensemaking analysis focused on hypotheses without explicit listing the supporting facts. Through those implicit mentions, six facts were covered but none were comprehensive. The \textbf{``step-by-step''} approach slightly improved LLM's performance, covering nine facts in total where five were comprehensive. See \autoref{fig:facts}.

\begin{figure}[h]
    \centering
    \includegraphics[width=0.75\linewidth]{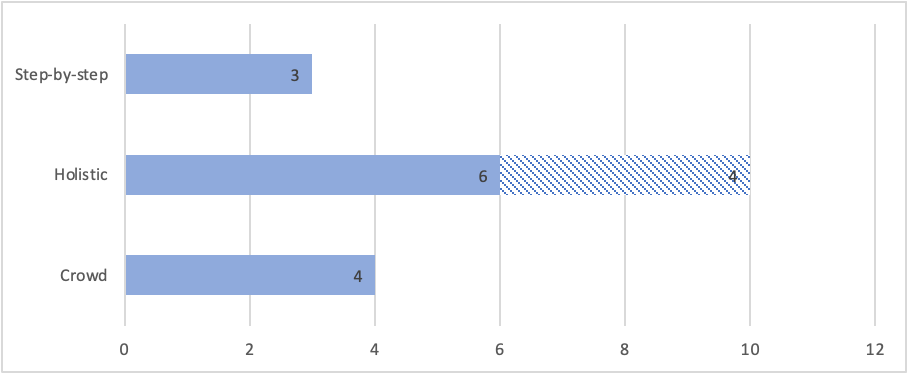}
    \caption{Efficacy in Hypotheses Development (shadowed parts are hypotheses that are nto comprehensively covered)}
    \label{fig:hypotheses}
\end{figure}

In terms of \textit{hypotheses development}, \textbf{crowdsourced} analyses covered 4 out of 15 predefined hypotheses, all are comprehensive. In a ``\textbf{holistic}'' scenario, LLMs generated hypotheses covering 10 hypotheses, of which 6 were comprehensive. With ``\textbf{step-by-step}'' approach, LLMs comprehensively covered 3 hypotheses. See \autoref{fig:hypotheses}.

\section{Implications}
LLMs demonstrate a strong capability in formulating hypotheses, especially by maintaining an overarching perspective of the data. While LLMs have the potential to outperform humans in hypothesis generation, this capability hinges on strategic prompting. Given the gaps in facts and hypotheses in initial findings, human intervention remains crucial for directing LLMs towards generating meaningful and accurate hypotheses.

Our results show that LLMs tend to merge the steps defined in the human sensemaking process~\cite{pirolli2005sensemaking}, extracting facts during document selection and forming hypotheses during fact extraction, challenging the structured approach of human sensemaking. 
The abstract nature of LLM-generated analysis underscores the need for human guidance to ensure the analysis is grounded in solid data, thereby avoiding potential inaccuracies.

It's intriguing to note that human analysis frequently showcases a higher level of comprehension in capturing the factual content of documents, while LLMs are more inclined towards aggressive hypothesis generation and data synthesis. This phenomenon resonates with the inherent cognitive constraints of humans and the formidable data processing capabilities of LLMs. Despite seeming counter-intuitive, a successful collaborative approach for sensemaking tasks could involve a hybrid model where humans lead the initial data gathering phase, leveraging their detailed understanding and commitment to accuracy. LLMs could then enhance this process by synthesizing and interpreting the collected data, utilizing their strengths in fast data processing and pattern recognition. Although further iterations and interaction are needed, this collaborative model would assign clear leadership roles in the information collection and synthesis stages.
Alternatively, LLMs could provide an initial broad analysis, which humans refine by probing the LLMs' logic and evidence through targeted inquiries, ensuring a more thorough and accurate exploration of the data.

\bibliographystyle{ACM-Reference-Format}
\bibliography{sample-base}










\end{document}